\documentclass[reprint,prl,aps]{revtex4-1}
\usepackage{graphicx}
\usepackage{color}
\usepackage{dcolumn}
\usepackage{bm}
\usepackage{amssymb}
\usepackage{color,amsmath}
\usepackage{soul,xcolor} 
\setstcolor{red} 
\usepackage{epstopdf}

\begin{document}

\title{Phonon scattering dominated electron transport in twisted bilayer graphene}

\author{Hryhoriy Polshyn$^{1*}$}
\author{Matthew Yankowitz$^{2*}$}
\author{Shaowen Chen$^{2,3}$}
\author{Yuxuan Zhang$^{1}$}
\author{K. Watanabe$^{4}$}
\author{T. Taniguchi$^{4}$}
\author{Cory R. Dean$^{2\dagger}$}
\author{Andrea F. Young$^{1\dagger}$}

\affiliation{$^{1}$Department of Physics, University of California, Santa Barbara, CA 93106}
\affiliation{$^{2}$Department of Physics, Columbia University, New York, NY, USA}
\affiliation{$^{3}$Department of Applied Physics and Applied Mathematics, Columbia University, New York, NY, USA}
\affiliation{$^{4}$National Institute for Materials Science, 1-1 Namiki, Tsukuba 305-0044, Japan}
\affiliation{$^{*}$These authors contributed equally to this work. $^{\dagger}$ afy2003@ucsb.edu (A.F.Y.); cd2478@columbia.edu (C.R.D.)}


\maketitle

\textbf{Twisted bilayer graphene (tBLG) has recently emerged as a platform for hosting correlated phenomena, owing to the exceptionally flat band dispersion that results  near interlayer twist angle $\theta\approx1.1^\circ$. At low temperature a variety of phases are observed that appear to be driven by electron interactions including insulating states, superconductivity, and magnetism~\cite{cao_correlated_2018,cao_unconventional_2018,yankowitz_tuning_2019,sharpe_emergent_2019}. Electrical transport in the high temperature regime has received less attention but is also highly anomalous, exhibiting gigantic resistance enhancement and non-monotonic temperature dependence. Here we report on the evolution of the scattering mechanisms in tBLG over a wide range of temperature and for twist angle varying from 0.75$^\circ$ - 2$^\circ$. We find that the resistivity, $\rho$, exhibits three distinct phenomenological regimes as a function of temperature, $T$. At low $T$ the response is dominated by correlation and disorder physics; at high $T$ by thermal activation to higher moir\'e subbands; and at intermediate temperatures $\rho$ varies linearly with $T$. The $T$-linear response is  much larger than in monolayer graphene~\cite{efetov_controlling_2010} for all measured twist angles, and increases by more than three orders of magnitude for $\theta$ near the flat-band condition. Our results point to the dominant role of electron-phonon scattering in twisted layer systems, with possible implications for the origin of the observed superconductivity.}

The electronic properties of twisted bilayer graphene are highly sensitive to the twist angle. At large angles, momentum mismatch between the rotated Brillouin zone corners effectively decouples the two layers. In contrast, at small angles interlayer-tunneling strongly hybridizes the layers, leading to a significantly reconstructed bandstructure. At the so-called ``magic angle'' ($\theta \approx 1.1^\circ$), a narrow, low energy, band appears~\cite{bistritzer_moire_2011,suarez_morell_flat_2010} in which the Fermi velocity becomes vanishingly small, and correlations play an important role~\cite{kim_tunable_2017}. Near this angle transport studies have revealed insulating states at band fillings corresponding to an integer number of electrons per moir\'e unit cell~\cite{cao_correlated_2018}, as well as superconducting states at a variety of partial band fillings~\cite{cao_unconventional_2018,yankowitz_tuning_2019}. However, despite intense theoretical effort there is little agreement upon the origin of the superconducting states, which have been proposed to arise from either an all-electronic mechanism mediated by magnetic fluctuations of the correlated insulating states~\cite{cao_unconventional_2018} or from a conventional phonon-mediated mechanism~\cite{wu_phonon_mediated_2018,lian_twisted_2018}.

\begin{figure*}[ht]
\includegraphics[width=6.9 in]{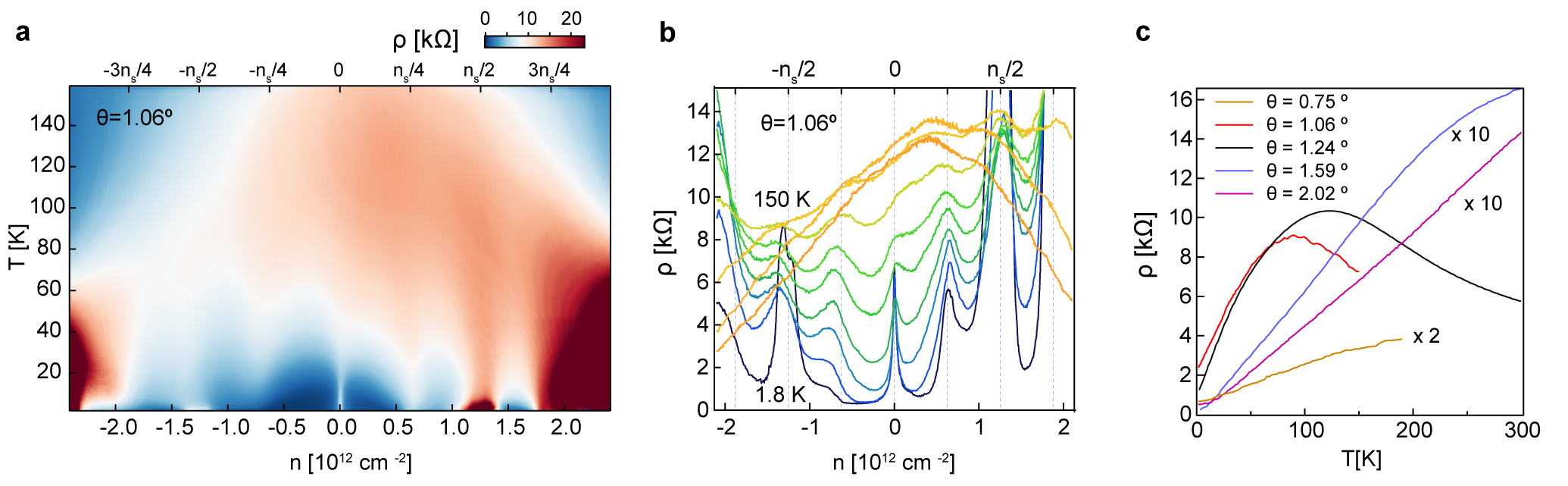}
\caption{\textbf{Temperature dependence of resistivity in small twist angle tBLG devices.}
\textbf{a}, Temperature dependence of $\rho$ in 1.06$^{\circ}$ tBLG device, which exhibits correlated states and superconductivity at the base temperature.
\textbf{b}, $\rho$ as a function of carrier density at selected temperatures in 1.06$^{\circ}$ tBLG device. $\rho$ traces were taken at temperatures of 1.8 (black), 6, 13, 20, 30, 40, 60, 90, 120 and 150~K (orange) respectively.
\textbf{c}, $\rho(T)$ measured in devices with different twist angles near $-n_s/2$ filling.
}
\label{fig:1}
\end{figure*}

For ordinary metals, the high-temperature resistivity is dominated by electron-phonon scattering and evolves linearly in temperature, with the slope, $d\rho/dT$, directly related to the electron-phonon coupling strength. At low temperatures this dependence can transition to a different power as phonon scattering becomes inefficient and scattering from electron-electron interactions or disorder plays a larger role. For many metals that become superconducting at low temperature, the dimensionless coupling constant, $\lambda_{tr}$, extracted from the $T$-linear region is found to correlate well with the coupling constant, $\lambda^*$, that determines the phonon-mediated superconducting transition temperature~\cite{allen_electron-phonon_2000}. High-temperature transport can therefore provide a sensitive probe of electron-phonon coupling strength and help elucidate its role in the correlated states that appear at low temperature. However, in many materials $\rho(T)$ can be complicated and therefore resistant to simple interpretation. In semiconductors both intraband scattering mechanisms and the thermal activation of carriers between bands contribute to the dependence of $\rho(T)$, requiring detailed theoretical analysis. In cuprate superconductors and other strongly correlated systems, $\rho(T)$ increases without saturation at high $T$, seemingly inconsistent with Fermi liquid theory~\cite{emery_superconductivity_1995}.
tBLG has large density of states in the flat band, small energy gaps to excited bands that can be easily bridged by thermal activation, and a low-temperature phase diagram in which strong electron-electron correlations likely play a role. Multiple contributions to $\rho(T)$ resembling those in metals, semiconductors, and strongly correlated systems may thus be expected, reflecting both the electronic structure and relevant scattering mechanisms.

Figs.~\ref{fig:1}a-b show $\rho(T)$ measured in a tBLG sample near the flat-band condition ($\theta$ = 1.06$^\circ$) for carrier densities spanning the lowest energy band (we label $\pm n_s$ as the density required to fill the lowest moir\'e subbands). Insulating response is observed at full band filling, characterized by high resistance peaks that diverge with decreasing temperature, consistent with previous measurements near this twist angle~\cite{kim_charge_2016,cao_superlattice-induced_2016,kim_tunable_2017}. We additionally observe correlated insulating phases or resistance peaks at several integer multiples of $n_s/4$, as well as superconducting states at a variety of partial band fillings, as previously reported elsewhere~\cite{cao_correlated_2018,cao_unconventional_2018,yankowitz_tuning_2019}. As the temperature is raised, the $\rho(T)$ at partial band filling exhibits a complex dependence. At most densities, $\rho(T)$ increases with increasing temperature, consistent with metallic behaviour (an exception to this general trend is observed at the correlated insulating states where $\rho(T)$ first decreases with $T$ but then eventually increases). At higher temperatures the resistance throughout the band saturates and then drops, exhibiting a negative $d\rho/dT$ with further temperature increase.

To disambiguate the role of electronic correlations in this temperature response we compare 7 tBLG devices with twist angles ranging from well below ($\sim$0.75$^\circ$) to well above ($\sim$2$^\circ$) the flat band condition. Fig.~\ref{fig:1}c shows $\rho(T)$ near $n=-n_s/2$ for 5 of these devices. As is evident in the plot, the behavior of the $\rho(T)$ is qualitatively similar between samples of all twist angles. Specifically, we identify three distinct temperature regimes marked by different behavior of $\rho(T)$. In the high-temperature regime, $\rho$ grows sub-linearly with increasing $T$, reaching a maximum at a temperature we define as $T_H$ before dropping again at the highest temperatures. Below this temperature, we find an intermediate regime where $\rho$ scales linearly with $T$. Finally, at the lowest temperatures $\rho(T)$ diverges from $T$-linear dependence. Depending on density and twist angle, the low temperature regime can be marked by resistivity saturation (most clearly observed in the 2.02$^\circ$ device in Fig.~\ref{fig:1}c), insulating, or superconducting behavior. These three regimes are not universally demarcated; they depend on both twist angle and carrier density.

\begin{figure*}[ht]
\includegraphics[width=6.9 in]{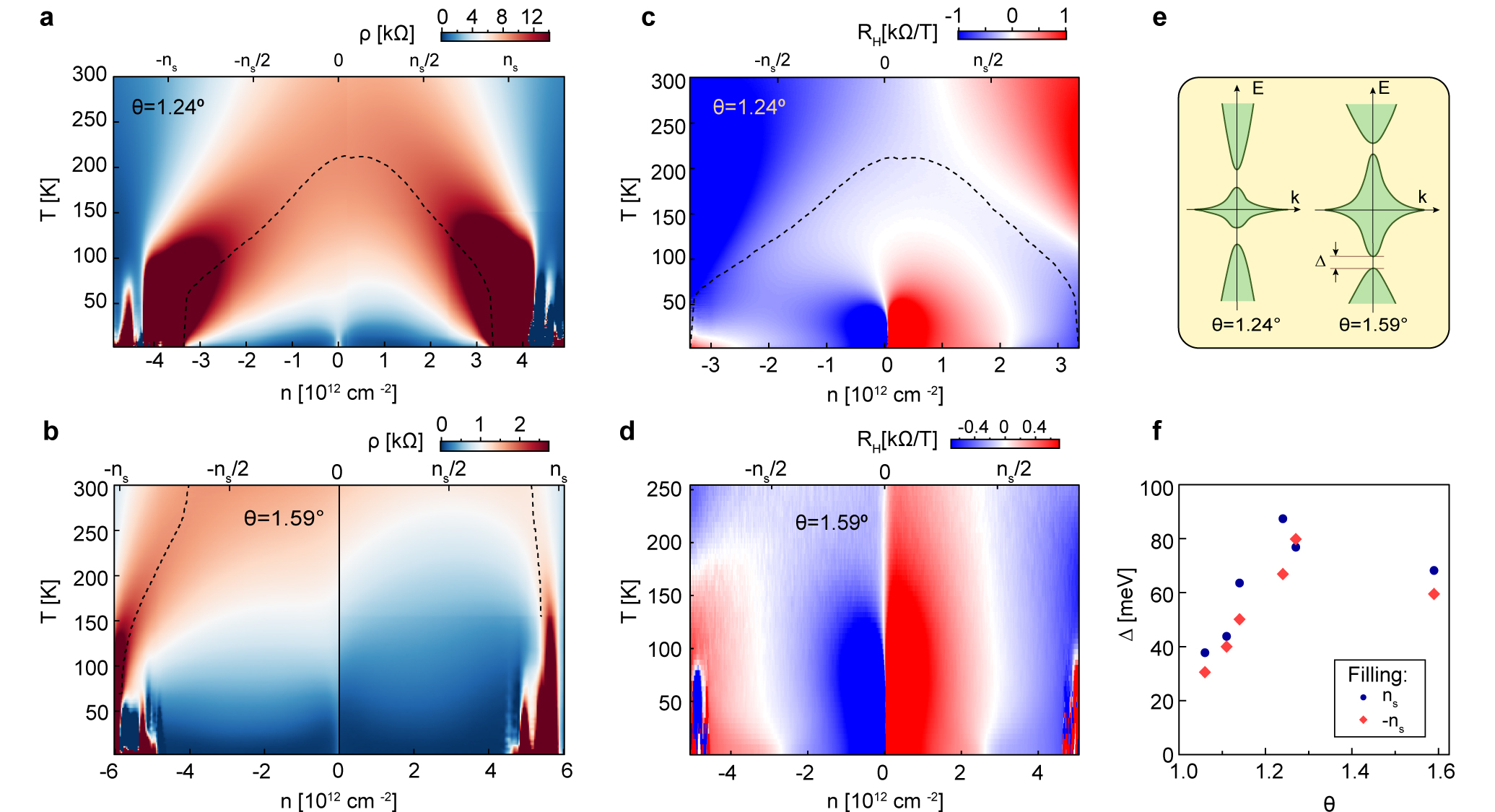}
 \caption{\textbf{High temperature behaviour of resistivity.}
Temperature dependence of $\rho$ in a 1.24$^{\circ}$ (\textbf{a}) and 1.59$^{\circ}$ (\textbf{b}) device. Dashed black line indicates the position of the peak values of $\rho(T)$ ($T_H$).
\textbf{c-d}, Hall coefficient $R_{H}(T)$, symmetrized at $B$ = 1 T, as a function of carrier density for the devices in (a) and (b), respectively. The dashed black line shown in (c) is reproduced from (a).
\textbf{e}, Schematic diagram of the band structure of $1.24^\circ$ and $1.59^\circ$ tBLG. The bandwidth of the low-energy moir\'e subbands grows with increasing twist angle.
\textbf{f}, Activation energy gaps of the $\pm n_s$ insulating states at various twist angles.
}
\label{fig:2}
\end{figure*}

We first discuss the high-temperature regime. Fig.~\ref{fig:2}a shows $\rho(T)$ for a 1.24$^\circ$ device in which no features associated with electronic correlations, such as partial band insulators or superconductors, are observed at low temperatures. Similar to the  1.06$^\circ$ device (Fig.~\ref{fig:1}a), we find that $\rho(T)$ first grows with increasing temperature up to a temperature $T_H$ (dashed curve in Fig. 2a), where it peaks before dropping as $T$ is raised further. We find that $T_H$ is largest near the charge neutrality point (CNP), and shrinks rapidly near $\pm n_s$. Fig.~\ref{fig:2}b shows a $\rho(T)$ map for a device with $\theta$ = 1.59$^\circ$, in which we observe a similar behavior but with larger $T_H$ at all carrier densities. The origin of the non-monotonicity of $\rho(T)$ becomes evident in the response of the Hall coefficient, $R_{H}$, plotted for the 1.24$^\circ$ device in Fig.~\ref{fig:2}c. At base temperature, the sign of $R_{H}$ switches at approximately $\pm n_s/2$, a featured associated with a Lifshitz transition at the van Hove singularity (vHs) in the tBLG band structure~\cite{kim_charge_2016,cao_superlattice-induced_2016}. However, we also find that $R_{H}$ evolves non-monotonically at fixed density as a function of $T$, changing sign with increasing temperature for densities $|n| \gtrsim n_s/2$. Overlaying the density-dependent $T_H$ curve on the $R_H$ data, it can be seen that $T_H$ closely tracks the temperature beyond which the magnitude of the Hall coefficient grows rapidly (Fig.~\ref{fig:2}c).

\begin{figure*}[ht]
\includegraphics[width=6.9 in]{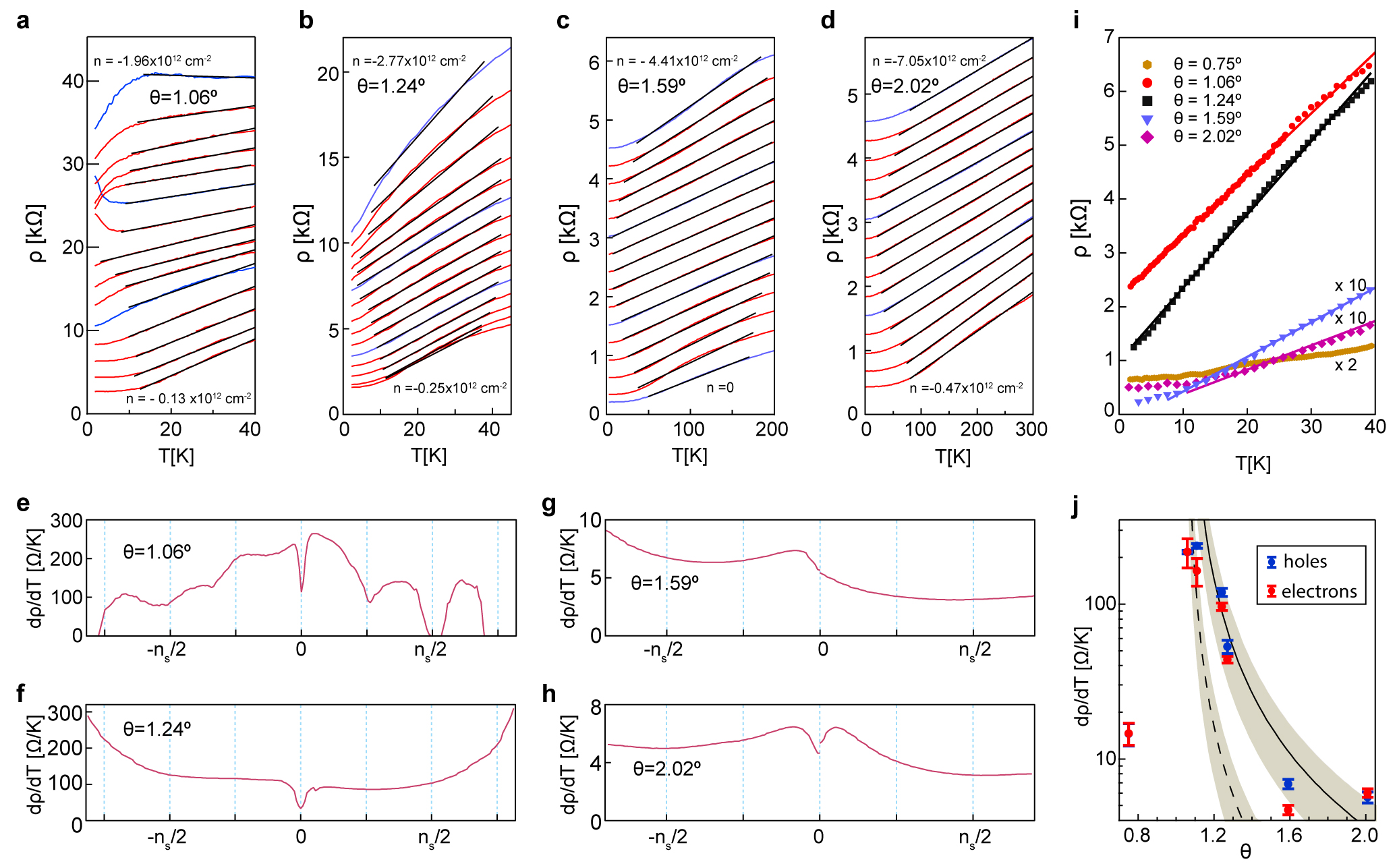}
\caption{\textbf{$T$-linear resistivity in devices with different twist angles.}
\textbf{a-d}, $\rho(T)$ curves at selected carrier densities for devices with twist angles of 1.06$^\circ$, 1.24$^\circ$, 1.59$^\circ$, and 2.02$^\circ$ respectively. Curves are offset by 2~k$\Omega$ in (a), by 600 $\Omega$ in (b), and by 300 $\Omega$ in (c)-(d). Solid black lines are guide for eye only
\textbf{e-h}, $d\rho/dT$ extracted in the $T$-linear regime corresponding to the devices in (a)-(d).
\textbf{i} $\rho(T)$ measured in devices with different twist angles near $-n_s/2$ filling.
\textbf{j}, $d\rho/dT$ of 7 tBLG devices averaged from $n = +(-)0.15$ to $+(-)0.5 \times 10^{12} \mathrm{cm}^{-2}$ for electrons (holes) in red (blue) markers. Dashed line shows $d\rho/ dT$ predicted by Eq. 1 with $D_A/v_s$ set to the values corresponding to monolayer graphene. The solid line shows the prediction with $D_A/v_s$ enhanced by a factor of 3. The shaded bands represent $\pm 50$\% error bar reflecting the effect of uncertainties in $D_A$, $v_{ph}$ and $v_F$.
}
\label{fig:3}
\end{figure*}

This observed behavior can be understood by considering the band structure of tBLG, which features low energy moir\'e subbands with small Fermi velocity. These bands are isolated from highly dispersive, higher energy subbands by sizable band gaps, which are predicted to arise in band structure calculations accounting for lattice relaxations~\cite{nam_lattice_2017}. We measure these gaps using thermal activation measurements (Fig. S7), finding that they are typically 30 - 90 meV, with the smallest gaps found in devices with twist angles near the magic angle (Fig.~\ref{fig:2}f). Fig.~\ref{fig:2}e shows a schematic of the tBLG band structure at two different twist angles. The correlation between $T_H$ and large changes in $R_H$ suggests that thermal activation of carriers to the dispersive bands plays a dominant role at high temperatures, with $T_H$ set by a combination of the bandwidth of the low-energy moir\'e subbands, the band gaps isolating the higher subbands, and the Fermi energy. Samples close to the magic angle have both smaller bandwidth and smaller band gaps, so that $T_H$ occurs at lower $T$ than in devices with larger twist angles.

$T_H$ thus provides a cutoff; for $T\lesssim T_H$, transport can be assumed to be restricted to the lowest electron- and hole-moir\'e subbands. Figs.~\ref{fig:3}a-d show $\rho(T)$ in this regime across a wide range of carrier densities between the CNP and $-3n_s/4$ for four devices with $\theta=$ 1.06$^\circ$, 1.24$^\circ$, 1.59$^\circ$, and 2.02$^\circ$, respectively. The black lines, which  are linear fits to $\rho(T)$,  are nearly parallel across all densities for all twist angles. A quantitative comparison of the linear slope versus density is shown in Figs.~\ref{fig:3}e-h.  We note that we restrict our analysis of $d\rho/dT$ to densities $|n|\lesssim 3n_s/4$ since at higher densities, the insulating states at $\pm n_s$ begin to significantly influence $\rho(T)$ even at low temperatures. For devices at twist angles far from the flat band condition,  $d\rho/dT$ is nearly constant with carrier density, although it is always slightly larger for holes than electrons. Flat band devices show somewhat more variation (Fig.~\ref{fig:3}e  and Fig.~S2), with step-like changes in $d\rho/dT$ around the quarter band fillings. Notably, signatures of commensuration with the lattice period, in the form of resistance peaks at integer multiples of $n_s/4$---including those showing no features in the low temperature limit---are evident even at elevated temperatures $T \approx$ 100 K, well above the onset of true insulating behavior (defined as $d\rho/dT<0$) at these fillings (Figs.~\ref{fig:1}a-b). Although we currently do not have a full understanding of these phenomena tied to the quarter band fillings, it appears they may share a common origin.

\begin{figure*}[ht]
\includegraphics[width=2.75 in]{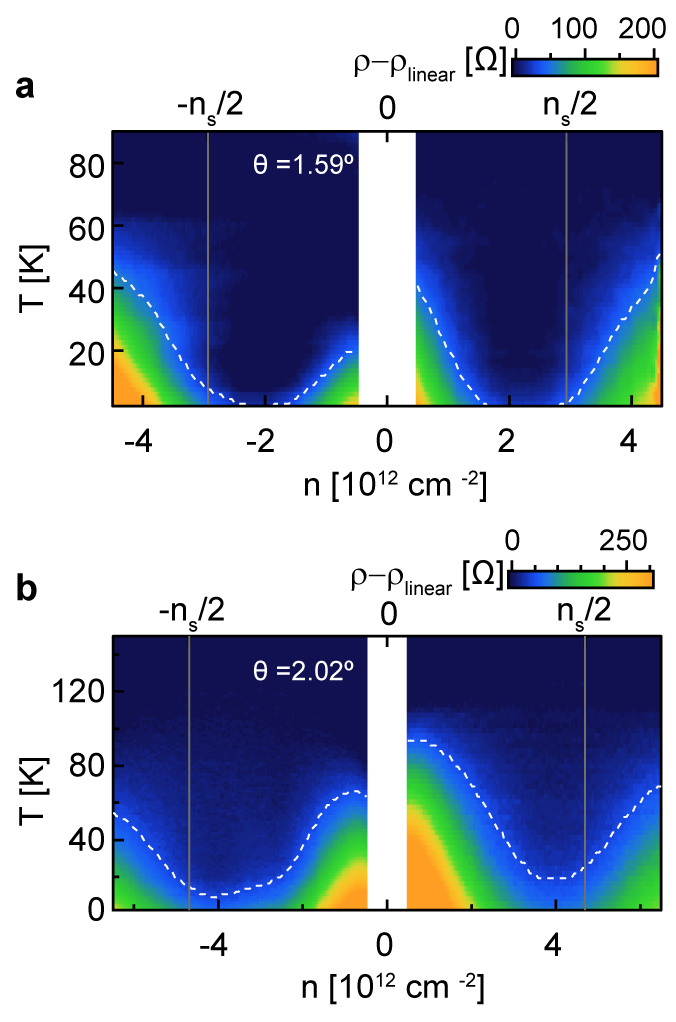}
\caption{\textbf{Deviation from $T$-linear resistivity at low temperature.}
Map of $\rho-\rho_{linear}$ at low temperatures in samples with twist angle 1.59$^\circ$ (\textbf{a}), and 2.02$^\circ$ (\textbf{b}). Dashed white line denotes the contour where $|\rho-\rho_{linear}| =50$~$\Omega$. For both twist angles, we find that $T_L$ is smallest near $\pm n_s/2$.
}
\label{fig:4}
\end{figure*}

While the $T$-linear behavior is qualitatively similar in all tBLG devices regardless of twist angle (Fig.~\ref{fig:3}i), the magnitude of $d\rho/dT$ varies strongly with $\theta$. Fig.~\ref{fig:3}j shows $d\rho/dT$ as a function of twist angle for the 7 devices presented in this study, averaged over carrier densities $|n|<0.5\times 10^{12}$~cm$^{-2}$. The magnitude of $d\rho/dT$ peaks for angles near the flat band condition ($\theta\approx 1.1^\circ$).
To understand this effect, we begin by noting that a similar (albeit much weaker) high temperature $T$-linear scaling of $\rho(T)$ is also observed in monolayer graphene (MLG), where it is attributed to scattering of electrons by thermally populated acoustic phonons \cite{chen_intrinsic_2008,dean_boron_2010,hwang_acoustic_2008,efetov_controlling_2010,hwang_acoustic_2008,efetov_controlling_2010}. Notably, $d\rho/dT$ ranges between 5-300 $\Omega$/K in all tBLG samples we study --- orders of magnitude larger than observed in MLG ($d\rho/dT\approx 0.1~\Omega$/K). Theoretically, resistivity in the $T$-linear regime  due to quasielastic scattering of acoustic phonon modes with classical occupation probability is given by~\cite{wu_phonon-induced_2018}
\begin{equation}
    \rho=\frac{\pi F D_A^2}{g e^2 \hbar \rho_m v_F^2 v_{ph}^2}k_B T.
    \label{eq:1}
\end{equation}
where $D_A$ is the deformation potential which describes the strength of electron-phonon coupling, $v_F$ and $v_{ph}$ are the Fermi and phonon velocities, respectively, and $\rho_m=7.6\times10^7$~kg/m$^2$. $g$ counts the number of electron flavors, with $g=4$ in MLG and $g=8$ in TBLG, while the form-factor $F$ accounts for differing electron-phonon matrix elements and is unity for MLG and $\sim$0.5 for tBLG. According to Eq. 1, a simple origin for the large and twist-angle dependent $d\rho/dT$ observed in tBLG could be attributed to strong renormalization of $v_F$ near the magic angle, which appears squared in the denominator of $\rho(T)$.

Eq. 1 contains three free parameters: $D_A$, $v_{ph}$, and $v_{F}$. To determine $v_F$, we note that most theoretical models of tBLG predict that $v_F$ is approximately linear in $\theta$ near the flat band angle~\cite{bistritzer_moire_2011}. To fix the slope of this line, we experimentally determine $v_F$ for two devices with twist angles of 1.24$^\circ$ and 1.59$^\circ$ from temperature dependent measurements of low-magnetic field quantum oscillations (see Fig. S6). From these measurements, we find that $v_F(\theta)=(0.37\pm0.12)\times(\theta-1.05^\circ)\times 10^6$~m/sec. The dashed line in Fig. 3j shows a comparison of this theory taking $D_A=25\pm5$ eV and $v_{ph}=2.1\times10^4$~m/sec, consistent with MLG and graphite literature~\cite{komatsu_interpretation_1964, efetov_controlling_2010}. The shaded region indicates the $\pm 50$\% bounds that follow from the experimental uncertainties in the monolayer $D_A$, $v_{ph}$, and in our determination of $v_F$.

While the theory is qualitatively consistent with the large increase in $d\rho/dT$ observed near the flat band condition, it quantitatively underestimates $d\rho/dT$ at all angles. However, quantitative agreement can be achieved if the tBLG deformation potential is somewhat larger, or the phonon velocity somewhat smaller, than in MLG. The solid line in Fig. 3j shows $d\rho/dT$ predicted with the model if $D_A/v_{ph}$ is increased by a factor of 3 relative to MLG. Despite the simplicity of the acoustic phonon scattering model, it is successful in explaining the large enhancement of $d\rho/dT$ arising from the reduced Fermi velocity in tBLG, including at twist angles where electron correlations are not thought to play a significant role (1.24$^\circ$-2.02$^\circ$).

Finally, we address the behavior of $\rho(T)$ in the low temperature limit. The fits to the intermediate temperature regime in Figs.~\ref{fig:3}a-d fail at low temperature, where $\rho(T)$ saturates, decreases more quickly, or, in the case of correlated insulating states near the flat band condition, reverses sign.  Due to the complications inherent to the low-temperature behavior of flat-band devices, we focus on the devices at 1.59$^\circ$ and 2.02$^\circ$ where no strong correlation-driven physics is observed at low temperature. Figs.~\ref{fig:4}a-b show the deviation from $T$-linear behavior, $\rho-\rho_{linear}$, in devices with $\theta = 1.59^\circ$ and 2.02$^\circ$, respectively, with $\rho_{linear}$ the linear fit to the the intermediate temperature regime. We find that the lower limit of the $T$-linear regime, $T_L$ (defined as the temperature where $|\rho-\rho_{linear}|=50$~$\Omega$) is minimal near $\pm n_s/2$ in both of these devices. Simple modelling of electron-phonon scattering does not reproduce this observation. Theory predicts a crossover from $T$-linear dependence to $\rho(T)\propto T^4$ power law in the low temperature limit~\cite{hwang_acoustic_2008,wu_phonon-induced_2018}, with the transition temperature, known as the Bloch-Gr\"uneisen temperature, $T_{BG}\propto \sqrt{n}$. In this picture, $T_L$ should be minimal near the CNP, contrasting with the experimental observation. We speculate that the behavior of $T_L$ may be related to the vHs in the band structure that occurs near half band filling. In flat band devices, $T_L$ is an erratic function of density, reflecting the complexity of the low-temperature phase diagram, but we note that it can persist to the base temperature of our measurement ($\sim$1.6 K) for certain values of $n$ (Fig.~\ref{fig:S11}).

In metals where the $T$-linear resistivity arises entirely from electron-phonon scattering, experimentally measured $d\rho/dT$ provides a direct measurement of the dimensionless electron-phonon coupling $\lambda_{tr}$ through its relation to the transport scattering time $\tau$ via $\hbar/\tau=2\pi\lambda  k_B T$.  Taking parameters for flat-band devices where $v_F\approx v_F^{MLG}/25$~\cite{cao_unconventional_2018}, $d\rho/dT \approx$ 200 $\Omega$/K, and electron density $n \approx 10^{12}$ cm$^{-2}$,
\begin{equation}
    \lambda_{tr}=\frac{e^2 v_F}{k_B}\sqrt{\frac{2n}{\pi}} \frac{d\rho}{dT}\approx 1.
\end{equation}
In many superconducting metals, $\lambda_{tr}$ agrees to within 20\% with the electron-phonon coupling $\lambda^*$ extracted from the superconducting transition temperature~\cite{allen_electron-phonon_2000}. Within a weak-coupling BCS theory $T_c$ is related to $\lambda^*$ by
\begin{equation}
    T_c\approx E_0 \exp\left(-1/\lambda^*\right),
    \label{BCS}
\end{equation}
with a similar relation holding for the strong coupling limit~\cite{mcmillan_transition_1968} of $\lambda^* \approx 1$. In conventional metals, the energy scale $E_0$ is a fraction of the Debye temperature. However, in a flat-band system, the electronic states that can participate in superconductivity are limited by the bandwidth ($W$), and it is natural to assume $E_0$ comparable to W $\approx$ 5 meV for tBLG near 1.1$^\circ$. Taking Eq.~\ref{BCS} with $\lambda^* \approx 1$, phonon-driven superconductivity may be expected \cite{wu_theory_2018,lian_twisted_2018,wu_phonon-induced_2018} at temperatures of order the bandwidth, well above experimentally reported transition temperatures~\cite{cao_unconventional_2018,yankowitz_tuning_2019}.

Eq.~\ref{BCS} is unlikely to accurately predict the superconducting transition temperature in tBLG; it does not quantitatively account for the finite bandwidth, and ignores the Coulomb repulsion that suppresses superconductivity in metals.  Other scattering mechanisms may also play a role in the high temperature transport~\cite{chung_transport_2018}. For example, Umklapp scattering processes~\cite{wallbank_excess_2019} and collisions with higher energy optical phonons are all expected to increase resistivity at finite temperature in tBLG. Any of these would lower the estimate of the dimensionless acoustic phonon coupling $\lambda_{tr}$, and thus of $T_c$. However, our estimate highlights the plausibility of the observed superconductivity arising from electron-phonon interactions~\cite{lian_twisted_2018,wu_phonon_mediated_2018}.

$T$-linear resistivity has also been observed in a variety of strongly correlated materials in which band theory is thought to fail~\cite{bruin_similarity_2013,emery_superconductivity_1995}, including many which become superconducting at low temperatures. In these materials, quantum critical fluctuations, rather than phonons, have been proposed as the dominant scattering mechanism leading to $T$-linear behavior~\cite{Zaanen2004}. Recently, similar `strange metal' physics has been invoked to explain $T$-linear resistivity observed in flat-band tBLG near densities at which correlated phases emerge at low temperature~\cite{cao_strange_2019}. However, our observation of ubiquitous $T$-linear behavior for \textit{all} twist angles and \textit{all} densities within the lowest moir\'e subband --- independent of the presence of correlated phases at low temperature --- points to a single, unified scattering mechanism unrelated to the electron correlation physics itself.

\section*{Methods}
tBLG devices are fabricated using the ``tear-and-stack'' method~\cite{kim_van_2016}. The devices consist of small-angle tBLG (0.75-2$^\circ$) encapsulated between flakes of hexagonal boron nitride (BN) with typical thickness of 25-50 nm. All devices have graphite top and bottom gates, except the devices with twist angles of 0.75$^\circ$, 1.06$^\circ$, and 1.11$^\circ$ which do not have a top gate. The use of graphite gates has previously been shown to result in very low charge inhomogeneity~\cite{Zibrov2017}. Samples are assembled using a standard dry transfer technique that utilizes a poly-propylene carbonate (PPC) film on top of a polydimethyl siloxane (PDMS) stamp~\cite{wang_one-dimensional_2013}. Completed heterostructures are transferred onto a Si/SiO$_2$ wafer. We avoid heating the sample above 180$^\circ$C during device processing to best preserve the intended twist angle between the two monolayer graphene flakes.

Transport measurements are conducted in a four-terminal geometry with typical ac current excitations of 1-10 nA using standard lock-in technique at 17.7 Hz. We gate the contact regions (which extend beyond the graphite bottom gate) to a high carrier density by applying a gate voltage to the Si (typically 5-50 V for SiO$_2$ thickness of $\sim$285 nm) to reduce the contact resistance. For devices with large twist angles, a combination of the top and bottom gates is required achieve $|n|>\pm n_s$. In such cases, $\rho(T)$ maps for electron- and hole-type doping are acquired separately (denoted by the black vertical line at $n=0$ in Fig.~\ref{fig:2}b), and consequentially the displacement field varies in addition to $n$. However, we do not observe any significant modification of $\rho(T)$ as a result.

The twist angle $\theta$ is determined from the values of charge carrier density at which the insulating states at $\pm n_s$ are observed, following $n_s = 8\theta^2/\sqrt{3}a^2$, where $a=0.246$~nm is the lattice constant of graphene. The values of $\pm n_s$ are determined from the sequence of quantum oscillations in a magnetic field which project to $\pm n_s$ (or $\pm n_s/2$ for devices near the magic angle).

\section*{acknowledgments}
The authors acknowledge discussions with L. Balents, T. Senthil, S. Das Sarma and F. Wu, and thank S. Das Sarma and F. Wu for sharing their unpublished theoretical results. Work at both Columbia and UCSB was funded by the Army Research Office under W911NF-17-1-0323. Sample device design and fabrication was partially supported by DoE Pro-QM EFRC (DE-SC0019443). AFY and CRD separately acknowledge the support of the David and Lucile Packard Foundation. K.W. and T.T. acknowledge support from the Elemental Strategy Initiative conducted by the MEXT, Japan and  the CREST (JPMJCR15F3), JST. A portion of this work was carried out at the Kavli Institute of Theoretical Physics, Santa Barbara, supported by the National Science Foundation under Grant No. NSF PHY-1748958.

\section*{Author contributions}
M.Y., S.C., H.P. and Y.Z. fabricated the devices. H.P., M.Y., and S.C. performed the measurements and analyzed the data. K.W. and T.T. grew the hBN crystals. C.R.D. and A.F.Y. advised on the experiments. The manuscript was written with input from all authors.

\section*{Competing interests}
The authors declare no competing interests.

%

\clearpage


\renewcommand{\thefigure}{S\arabic{figure}}
\renewcommand{\thesubsection}{S\arabic{subsection}}
\setcounter{secnumdepth}{2}
\renewcommand{\theequation}{S\arabic{equation}}
\renewcommand{\thetable}{S\arabic{table}}
\setcounter{figure}{0}
\setcounter{equation}{0}

\section*{Supplementary Information}

\subsection{Properties of magic angle devices}

\begin{figure*}[h]
\includegraphics[width=6.9 in]{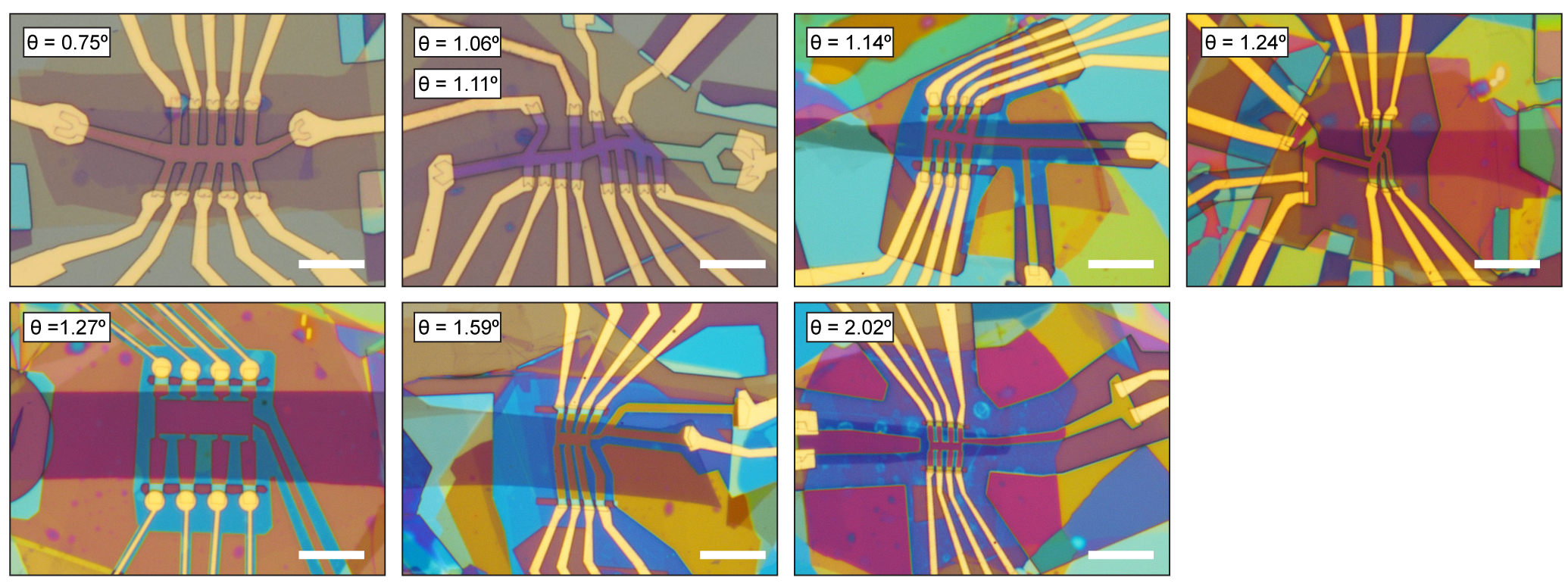}
\caption{\textbf{Optical microscope images of tBLG devices.} The twist angles of the sections of the devices that were measured  are shown in the top-left corners of the images.
All scale bars are 10~$\mathrm{\mu}$m.
}
\label{fig:S0}
\end{figure*}

We examine three tBLG devices near the magic angle, with twist angles of 1.06$^\circ$, 1.11$^\circ$ and 1.14$^\circ$. Optical images of these and other devices are shown in Fig.~\ref{fig:S0}. The low temperature characterization of superconducting and correlated insulating states for these devices was previously reported in Ref.~\cite{yankowitz_tuning_2019}. The 1.06$^\circ$ and 1.11$^\circ$ devices are independent sections of a large multi-contact device (D5 in Ref.~\cite{yankowitz_tuning_2019}) in which we observe superconductivity with $T_c \approx$ 0.9~K. We note that we report the values of resistance $R$ instead of resistivity $\rho$ for the 1.14$^\circ$ device (D1 in Ref.~\cite{yankowitz_tuning_2019}) because this devices was measured in psuedo-Van der Pauw geometry that precludes an exact conversion from $R$ to $\rho$.

\begin{figure*}[h]
\includegraphics[width=6.9 in]{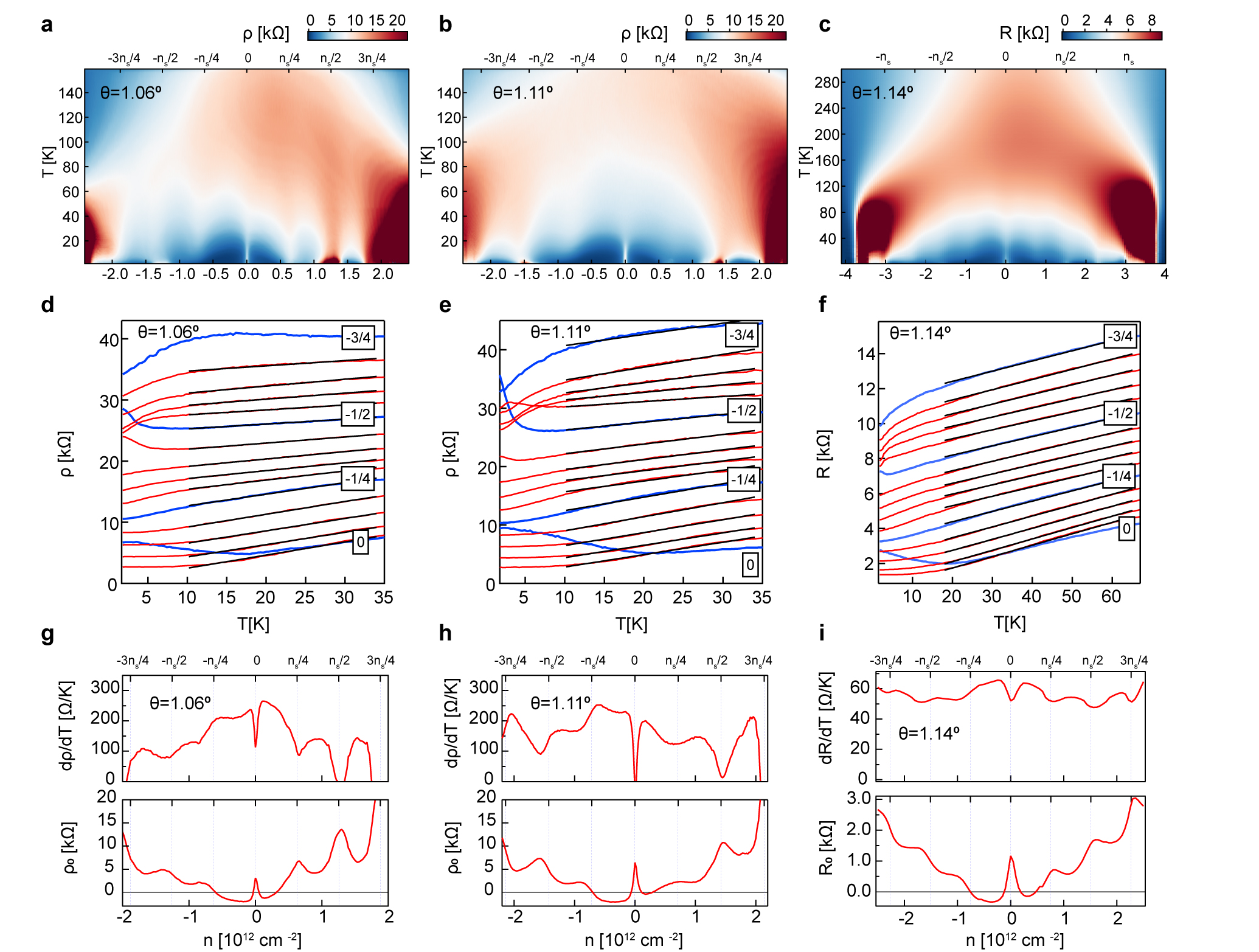}
\caption{\textbf{Measurements of resistivity in magic angle tBLG devices.}
\textbf{a}-\textbf{c}, $\rho(T)$ maps for devices with twist angles of 1.06$^\circ$ (a), 1.11$^\circ$ (b) and 1.14$^\circ$ (c, reported in $R(T)$).
\textbf{d}-\textbf{f}, $\rho(T)$ curves at selected carrier densities for the same devices as in (a)-(c). Blue traces are taken at fractional fillings of the band: $-3n_s/4$, $-n_s/2$, $-n_s/4$ and 0. All devices exhibit $T$-linear behaviour at nearly all band fillings to temperatures $T \lesssim$ 10~K. Black lines are linear fits (typically terminated at $T > T_L$).
\textbf{g}-\textbf{i}, Corresponding $d\rho/dT$ (top) and residual resistivity $\rho_0$ (bottom) extracted in the $T$-linear regime.}
\label{fig:S3}
\end{figure*}

Fig.~\ref{fig:S3}a-c show $\rho(T)$ maps ($R(T)$ for the 1.14$^\circ$ device). All three devices exhibit similar behaviour. At low temperatures we observe insulating states at $\pm n_s$, as well as resistive or insulating states at partial band filling. At high temperatures, the devices are most resistive closer to the CNP. Figs.~\ref{fig:S3}d-f show $\rho(T)$ curves taken at selected partial band fillings between 0 and $-3n_s/4$. Over the intermediate temperature range, we observe $\rho(T) \propto T$ for all $n$. The top panels of Figs.~\ref{fig:S3}d-f show the corresponding $d\rho/dT$, while the bottom panels show the residual resistivity, $\rho_0$, defined as the intercept of linear fits at $T=0$. Both $d\rho/dT$ and $\rho_0$ show sharp step-like jumps every time the carrier density crosses a multiple of $n_s/4$ filling of the band. In general, $d\rho/dT$ decreases and $\rho_0$ increases with almost every such step away from CNP in both directions. Presently, we do not have a complete understanding of the origin of this effect.

\begin{figure*}[h]
\includegraphics[width=4.75 in]{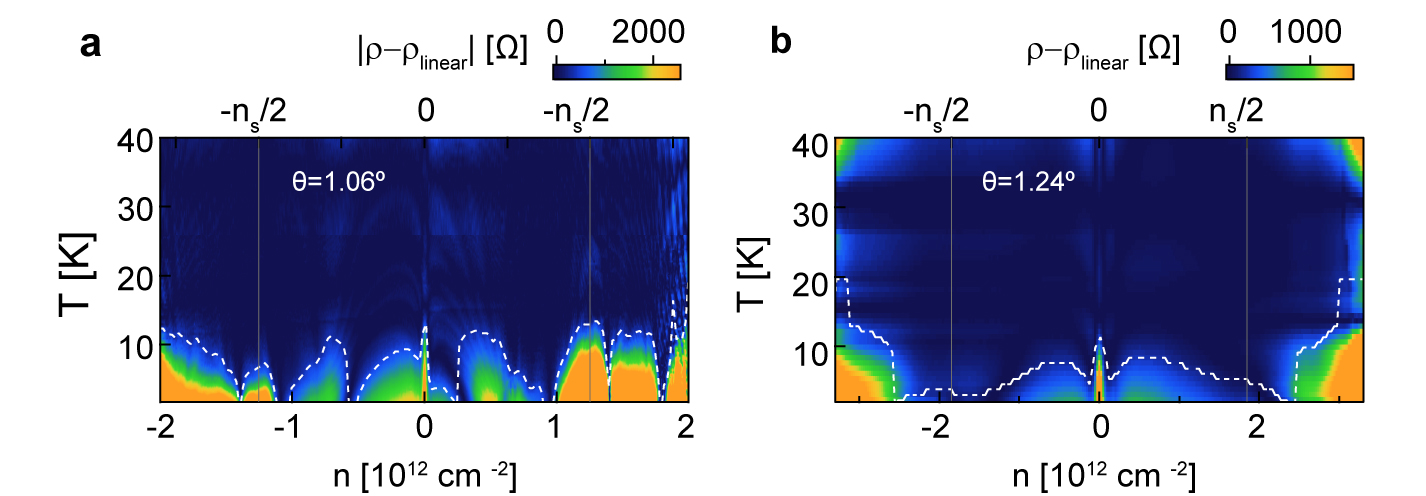}
\caption{\textbf{Deviation from $T$-linear behaviour at low temperatures.}
$\rho-\rho_{linear}$ at low temperatures in the (\textbf{a}) 1.06$^\circ$ sample and (\textbf{b}) 1.24$^\circ$ sample. For 1.06$^\circ$ sample the absolute value of the deviation is plotted. Dashed lines indicate contours where $|\rho-\rho_{linear}|\approx 500$~$\Omega$ and $300$~$\Omega$ for  (\textbf{a}) and (\textbf{b}) respectively. Both devices exhibit $T$-linear behavior to lowest temperatures near $\pm n_s/2$.
}
\label{fig:S11}
\end{figure*}

Fig.~\ref{fig:S11}a shows the deviation of the resistivity  $\rho-\rho_{linear}$ from $T$-linear behaviour in the 1.06$^\circ$ device at low temperatures. The deviation is shown as the difference between $\rho$ and an interpolation value of $\rho_{linear}$, obtained from fitting the $T$-linear behaviour at higher temperatures. The low temperature $\rho(T)$ shows a complicated dependence on $n$, and changes qualitatively across each quarter-filling of the band. The $T$-linear behavior persists to the lowest temperatures near the half-filling of the band ($\sim0.42 n_s)$. We observe qualitatively similar behavior in all our devices, with a comparable plots for the 1.59$^\circ$ device and the 2.02$^\circ$ device shown in Fig.~4 of the main text.

\subsection{Properties of non-magic angle devices}

\begin{figure*}[h]
\includegraphics[width=4.75 in]{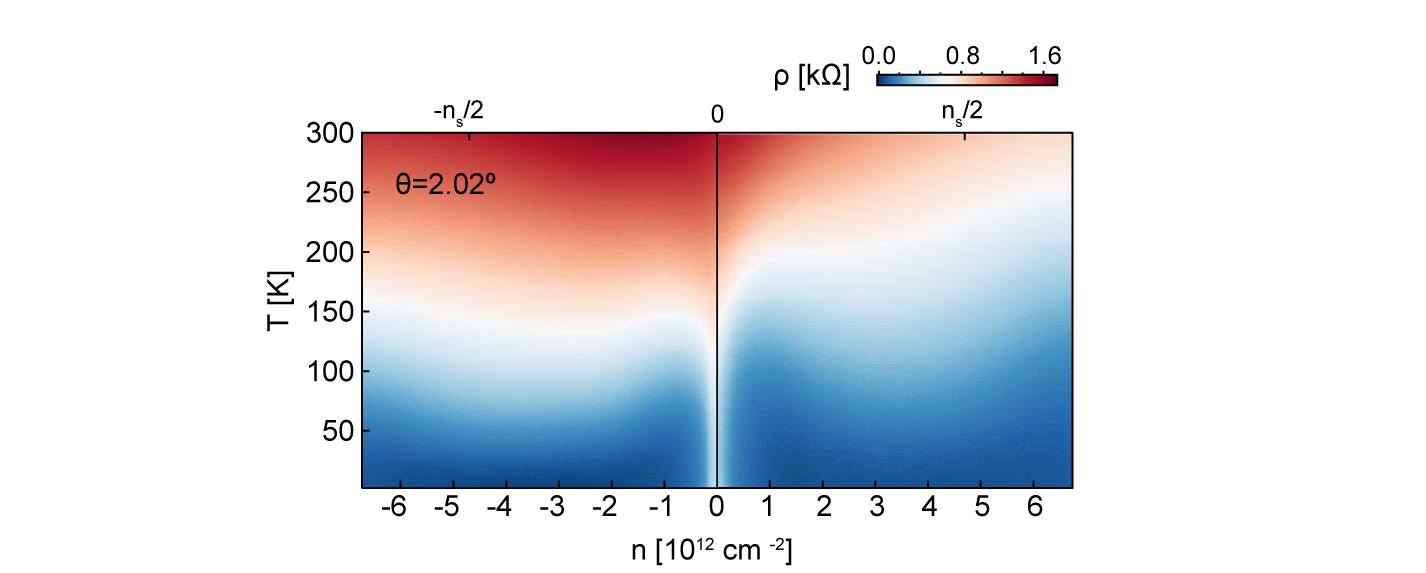}
\caption{\textbf{Temperature dependence of a 2.02$^\circ$ device.}
$\rho(T)$ map for the 2.02$^\circ$ device.
}
\label{fig:S9}
\end{figure*}

Figure~\ref{fig:S9} shows the $\rho(T)$ map for the 2.02$^\circ$ device. Notably, over the accessible range of $T$ and $n$ we no longer observe the ``high-temperature'' regime at this angle (i.e. $T_H >$ 300 K at all accessible $n$).

\begin{figure*}[h]
\includegraphics[width=6.9 in]{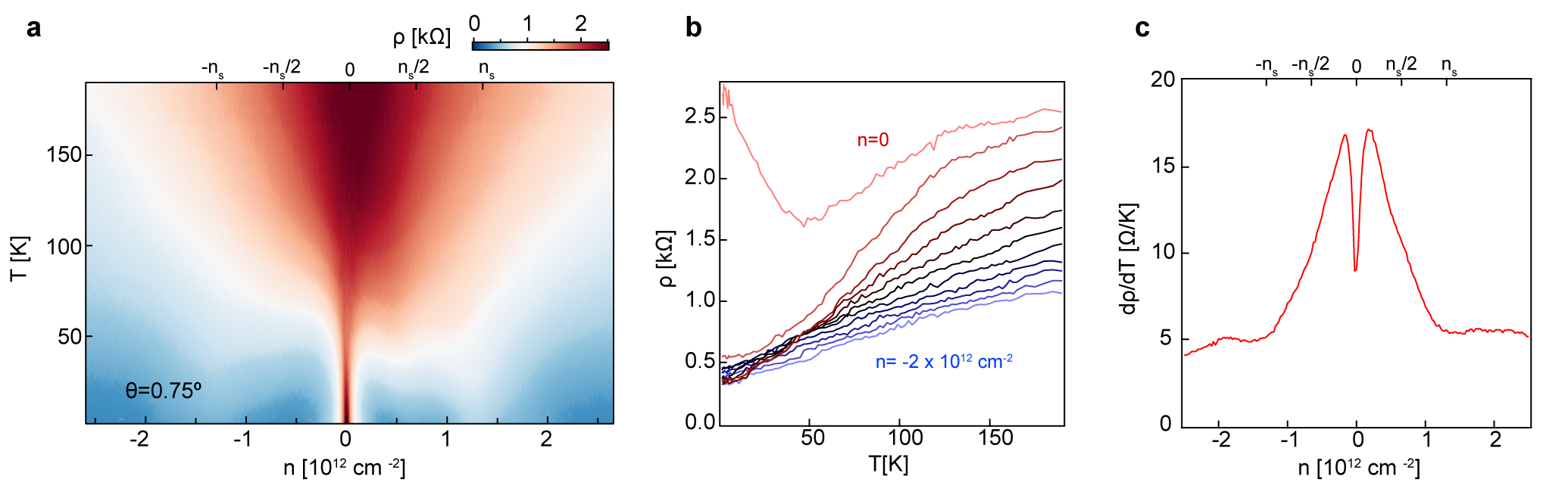}
\caption{\textbf{Temperature dependence of a 0.75$^\circ$ device.}
\textbf{a}, $\rho(T)$ map.
\textbf{b}, $R(T)$ at selected $n$ between 0 and $-2\times 10^{12} \mathrm{cm}^{-2}$.
\textbf{c} $d\rho/dT$ as a function of $n$, taken over the range of temperature in which $\rho(T)$ is linear.
}
\label{fig:S5}
\end{figure*}

Fig.~\ref{fig:S5}a shows a $\rho(T)$ map for the 0.75$^\circ$ device. This device exhibits considerably different phenomenology than devices with larger twist angles ($\theta > 1^\circ$). At the lowest measured temperature, there is a strong peak in $\rho$ at the CNP, but only very weak peaks at $\pm n_s$. We observe a $T$-linear dependence in $\rho$ for $|n|>2\times10^{12}\mathrm{cm}^2$, with $d\rho/dT \approx$ 5 $\Omega$/K (Figs.~\ref{fig:S5}b-c). There is a more complicated $\rho(T)$ dependence at lower $n$, which is likely the result of dominant thermal activation processes arising from the small bandwidth and absence of band gaps isolating the lowest moir\'e subbands.

\subsection{Measurement of $m^*$ and $v_F$}

\begin{figure*}[h]
\includegraphics[]{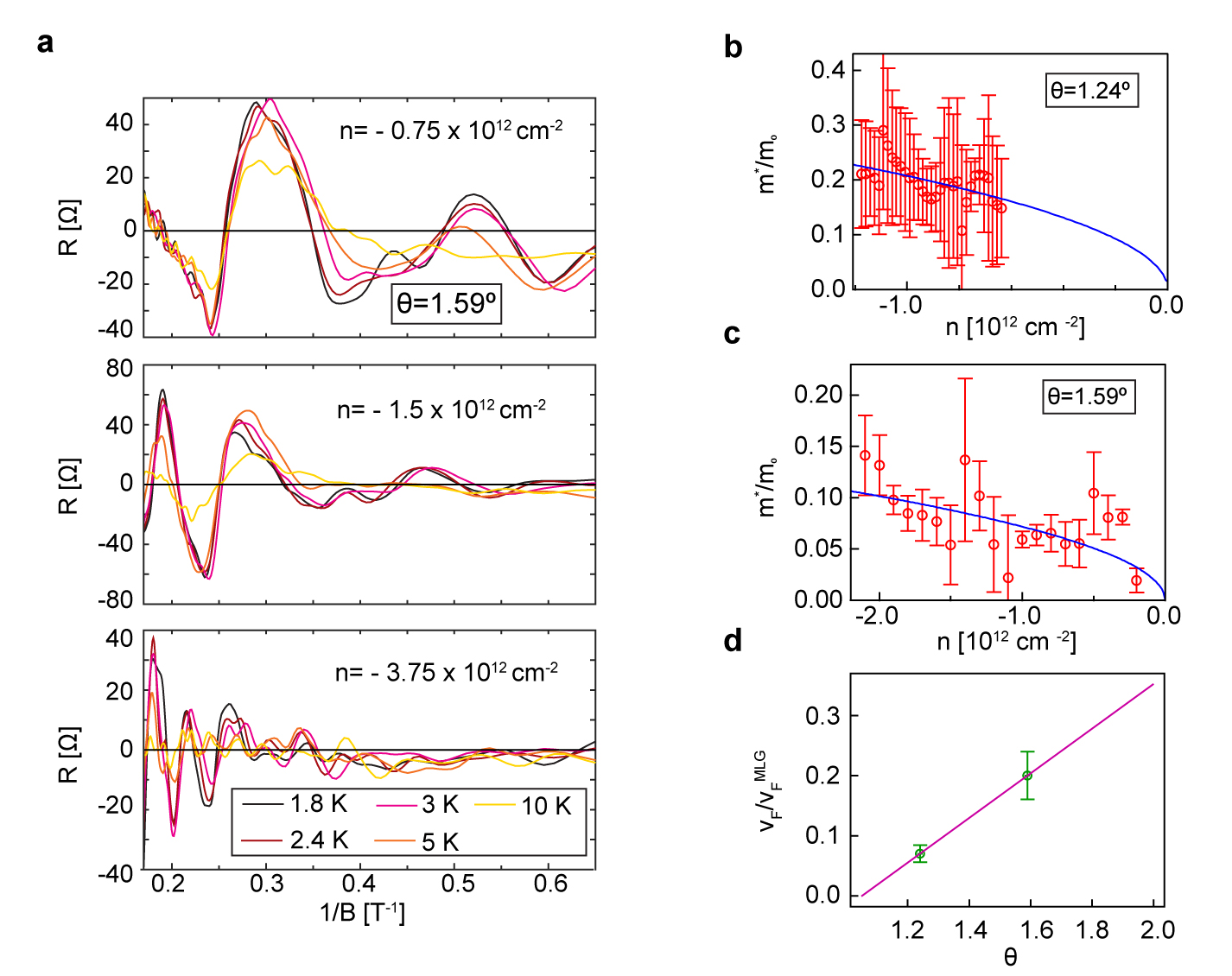}
\caption{\textbf{Measurements of $m^*$ in tBLG devices.}
\textbf{a}, Magnetoresistance measured at several temperatures for $n=-0.75$ (top), $-1.5$ (middle) and $-3.75\times 10^{12} \text{cm}^{-2}$ (bottom) in the 1.59$^\circ$ device.
\textbf{b}-{c} Extracted $m^*$ for the 1.24$^\circ$ and 1.59$^\circ$ devices, respectively. $v_F$ is extracted by fitting to $m^*=\sqrt{\hbar^2 n /8 \pi v_F^2}$ (blue curves).
\textbf{d}, $v_F$ as a function of $\theta$. The blue line shows a linearly extrapolation using the two measured values of $v_F$.
}
\label{fig:S2}
\end{figure*}

We measure the effective mass as a function of $n$ in the devices with twist angles of 1.24$^\circ$ and 1.59$^\circ$ by fitting the temperature dependence of the amplitude of Shubnikov–-de Haas quantum oscillations using the Lifshitz-Kosevich formula:
\begin{equation}
    \Delta R \propto \frac{\chi}{\sinh{\chi}},
    \label{eq:S1}
\end{equation}
where $\chi=2 \pi^2 k T m^*/(\hbar e B)$.
In order to isolate the amplitude of quantum oscillations $\Delta R$ from the other contributions to resistance, we first remove a background from the $R(H)$ curves by subtracting a polynomial fit to the data. The resulting $\Delta R$ curves at several $n$ are shown in Fig.~\ref{fig:S2}a.

For each $n$, we extract the amplitude of the most prominent peak at various $T$ and fit to Eq.~\ref{eq:S1}. The resulting values of $m^*(n)$ are shown in Fig.~\ref{fig:S2}b-c. We extract $v_F$ by fitting $m^*(n)$ to a Dirac dispersion expression $m^*=\sqrt{h^2 n /8 \pi v_F^2}$. We find a Fermi velocity of $v_F=7.0\pm0.5\times10^4$~m/s for the 1.24$^\circ$ device and $v_F=2.0\pm0.2\times10^5$~m/s for the 1.59$^\circ$ device. While the random error of $v_F$ is moderate ($\sim10$\%), we anticipate a considerable additional systematic error due to the background subtraction procedure used to isolate $\Delta R$. We estimate the total error for $v_F$ measurement to be $\approx20$~\%. In order to estimate the Fermi velocity for devices with other twist angles we linearly extrapolate the measured values of $v_F$, as shown in Fig.~\ref{fig:S2}d. This yields the following dependence on twist angle:
\begin{equation}
    v_F(\theta)=(0.37\pm0.12)\times(\theta-1.05^\circ)\times 10^6\text{m/sec}.
\end{equation}

\begin{figure*}[h]
\includegraphics[width=4.75 in]{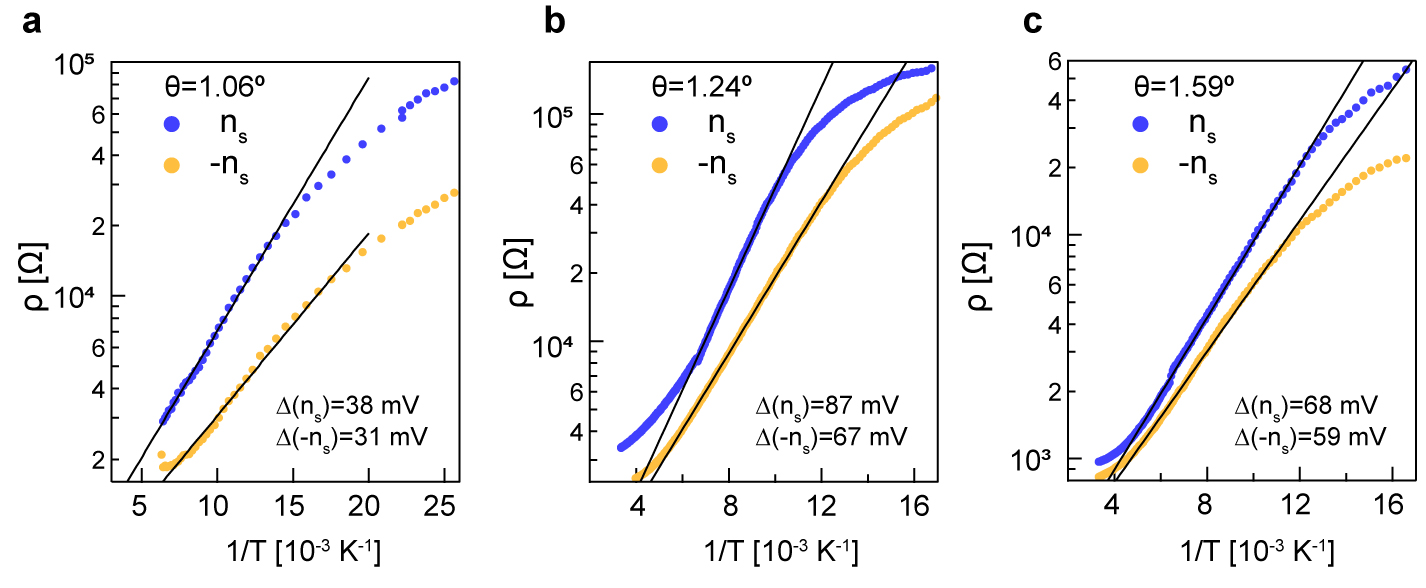}
\caption{\textbf{Measurements of band gaps at full band filling.}
$\rho(T)$ at $\pm n_s$ shown on an Arrhenius plot for \textbf{a} the 1.06$^\circ$ device, \textbf{b} the 1.24$^\circ$, and \textbf{c} the 1.59$^\circ$ device. $\Delta$ is extracted from a fit to the Arrhenius equation $\rho\propto \exp{\left[-\frac{\Delta}{2k_B T}\right]}$ over the thermally-activated regime.
}
\label{fig:S7}
\end{figure*}

\end{document}